\documentclass[twocolumn,twocolappendix]{aastex7}

\usepackage[T1]{fontenc}
\usepackage{graphicx}
\usepackage{amsmath}
\usepackage{placeins}
\usepackage{xcolor}
\definecolor{BrickRed}{rgb}{0.8,0.1,0.1}

\newcommand{\figfile}[2][]{%
  \IfFileExists{#2}{\includegraphics[#1]{#2}}{%
    \fbox{\parbox[c][2.2in][c]{\linewidth}{\centering Missing figure file:\\\texttt{#2}}}%
  }%
}

\newcommand{\ci}{[\ion{C}{1}]}
\newcommand{\ciitwo}{[\ion{C}{1}]($^3P_2\!\rightarrow{}^3P_1$)}
\newcommand{\co}{CO}
\newcommand{\Lsun}{L_\odot}
\newcommand{\Msun}{M_\odot}
\newcommand{\kms}{\ensuremath{\mathrm{km\,s^{-1}}}}

\shorttitle{ALMA High-J CO Spectroscopy II}
\shortauthors{Tadaki}

\begin{document}

\title{ALMA High-\emph{J} CO Spectroscopy of High-Redshift Galaxies. II. 0\farcs03 Resolution CO Kinematics Reveal Super-Eddington Accretion in a Dust-Obscured Galaxy at \emph{z}=3.111}

\author{Ken-ichi Tadaki}
\email{tadaki@hgu.jp}
\affiliation{Faculty of Engineering, Hokkai-Gakuen University, Toyohira-ku, Sapporo 062-8605, Japan}

\begin{abstract}
We present ultra-high-resolution ($0\farcs03 \approx 230$~pc) Atacama Large Millimeter/submillimeter Array (ALMA) observations of the hyperluminous dust-obscured galaxy W2305$-$0039 at $z = 3.111$, targeting the CO $J = 7$--$6$ and $J = 11$--$10$ lines.
The CO\,(11--10) emission is extremely compact and exhibits anomalously high excitation relative to CO\,(7--6) within the central $\lesssim$500~pc.
X-ray-dominated region models successfully reproduce this excitation, providing strong evidence for intense X-ray irradiation by a deeply obscured active galactic nucleus (AGN), while photodissociation-region models fail to match the observed ratio.
Forward modeling of the nuclear CO(11--10) position--velocity diagram yields a dynamical black-hole mass of $\log(M_{\rm BH}/M_{\odot}) = 8.3^{+0.7}_{-0.6}$ and an intrinsic gas velocity dispersion of $\sigma_{\rm gas} = 277^{+16}_{-14}$~\kms.
Combined with the AGN luminosity from infrared spectral energy distribution decomposition, these measurements imply a highly super-Eddington accretion state with $\lambda_{\rm Edd} \gtrsim 4$.
Our results provide dynamical evidence that the most rapid phases of black-hole growth can occur within a compact, heavily obscured nuclear region.
Extending ALMA beyond its current 16~km maximum baselines will be essential for pushing such dynamical measurements to tens-of-parsec scales and resolving the black-hole sphere of influence in massive galaxies at $z \gtrsim 6$.
\end{abstract}


\section{Introduction}
\label{sec:intro}

Luminous quasars at $z\gtrsim6$ already host black holes with masses of order $10^9\,\Msun$ less than a billion years after the Big Bang \citep[e.g.,][]{Mortlock2011,Wu2015,Matsuoka2018,Marshall2023,Yang2023,Loiacono2024}.
Explaining how these objects assembled so quickly remains a central problem in galaxy evolution and early structure formation.
The difficulty is not only to create sufficiently massive initial seed black holes, but also to sustain rapid growth over cosmic timescales that are short compared to typical black-hole e-folding times at the Eddington limit \citep[e.g., reviews by][]{Volonteri2010,Inayoshi2020}.
A widely discussed route to rapid assembly involves episodes of super-Eddington accretion, which can shorten the instantaneous mass-doubling time by factors of several to tens relative to Eddington-limited growth \citep[e.g.,][]{Jiang2014,Sadowski2014,Volonteri2015}.

Super-Eddington growth is expected to occur in gas-rich environments and to be accompanied by strong dust obscuration.
In theoretical frameworks coupling black-hole fueling and feedback to galaxy growth, the most intense accretion episodes are frequently those in which the nucleus is most heavily obscured \citep[e.g.,][]{Hopkins2006,HickoxAlexander2018}.
Observationally, many of the most rapidly growing black holes inferred at lower redshift are obscured, including Compton-thick active galactic nuclei (AGN) \citep[e.g.,][]{Ricci2015}.
At high redshift, however, heavy obscuration makes the most rapidly growing black holes difficult to identify using rest-frame optical/UV diagnostics alone, and also complicates black-hole mass estimates based on broad emission lines.

Among the most promising candidates for such obscured, rapid black-hole growth are the luminous galaxies selected in the mid-infrared with \emph{WISE} via the ``W1W2-dropout'' criterion, i.e., sources that are strongly detected at 12 and 22~\micron\ but weakly or not detected at 3.4 and 4.6~\micron, implying extreme mid-infrared obscuration \citep{Wright2010,Eisenhardt2012,Wu2012}.
These objects, commonly known as hot dust-obscured galaxies (Hot DOGs), constitute a very rare class of hyperluminous dust-obscured systems \citep[e.g.,][]{Assef2015}.
Hot DOGs exhibit spectral energy distributions (SEDs) dominated by a luminous mid-infrared component (rest-frame $1$--$20~\micron$), with an additional far-infrared component at $\gtrsim20~\micron$ that can include dust heated by star formation.
Their mid-infrared SEDs typically peak at rest-frame $4$--$10~\micron$ and show an excess around $\sim6~\micron$, consistent with a hot dust component of characteristic temperature $T_{\rm d}\sim450$~K, which strongly suggests that their enormous luminosities are predominantly powered by a heavily dust-obscured AGN \citep{Tsai2015}.
In Paper~I of this series \citep{2026arXiv260223521T}, we compiled ALMA-archive-based high-$J$ CO spectral line energy distributions for $z>3$ galaxies and found tentative evidence for enhanced excitation in Hot DOGs, consistent with AGN-related heating in deeply obscured nuclei.
This picture is reinforced by multi-wavelength SED decomposition, which indicates that Hot DOGs host extremely luminous, heavily obscured AGN at $z\sim2$--4 \citep[e.g.,][]{Sun2024}, and by several studies reporting Eddington ratios near or above unity based on single-epoch virial black-hole masses from rest-frame UV spectroscopy \citep[e.g.,][]{Tsai2018,Banados2021,2024ApJ...971...40L,Luo2025}.
However, virial estimates can be biased by non-virial components and uncertain extinction corrections in obscured systems \citep[e.g.,][]{Coatman2017,Ricci2022}.
Dynamical black-hole mass constraints that are insensitive to dust are therefore essential for testing whether the most extreme obscured systems truly accrete above the classical Eddington limit.

ALMA enables us to probe the inner few hundred parsecs of luminous dusty galaxies at $z\sim3$--6, where the black-hole gravitational potential and the circumnuclear interstellar medium (ISM) can both leave observable signatures in far-infrared and submillimeter emission.
In luminous quasars, recent ultra--high-resolution ALMA observations of the [\ion{C}{2}] $158~\micron$ fine-structure line have begun to resolve the host-galaxy ISM on sub-kiloparsec scales, enabling detailed studies of galaxy-wide rotation and feedback signatures \citep[e.g.,][]{Venemans2019,Walter2022,Neeleman2023,2025AA...695L..18M}.
Because [\ion{C}{2}] is typically among the brightest far-infrared lines on galaxy scales, it provides a highly efficient tracer of global host-galaxy dynamics and outflows.
However, [\ion{C}{2}] emission frequently extends over kiloparsec scales, and in the longest-baseline configurations, limited surface-brightness sensitivity and spatial filtering of low spatial frequencies can resolve out a substantial fraction of this extended flux \citep{Meyer2025}.
As a result, even deep integrations may provide limited leverage on the gas motions within the central few hundred parsecs that most directly probe the black-hole potential.
In contrast, high-$J$ \co\ emission preferentially traces warm and dense molecular gas and can be markedly more compact, remaining detectable at $\lesssim0\farcs05$ resolution \citep{Tadaki2025}.
High angular resolution therefore enables kinematic modeling of compact molecular disks, providing a dust-insensitive route to constrain black-hole masses in obscured nuclei \citep[e.g.,][]{Davis2013,Liao2025}.

Highly excited \co\ transitions provide sensitive diagnostics of the dominant heating mechanism.
Very high-$J$ \co\ emission and extreme high-$J$/mid-$J$ line ratios are difficult to explain with photodissociation regions (PDRs) alone and can instead indicate X-ray--dominated regions (XDRs) powered by an AGN \citep[e.g.,][]{Werf2010,Gallerani2014,Vallini2019,Esposito2024,Tadaki2025}.

In this paper, we present ultra--high-resolution ALMA observations of the Hot DOG W2305$-$0039 at $z=3.111$.
Our goals are (i) to quantify the radial structure of the warm molecular gas, (ii) to diagnose the dominant excitation mechanism in the central kiloparsec, and (iii) to constrain the black-hole mass dynamically from the nuclear \co\ kinematics and thereby infer the accretion state of the obscured AGN.
Throughout this paper, we assume a flat $\Lambda$CDM cosmology with $H_{0}=70~{\rm km~s^{-1}~Mpc^{-1}}$ and $\Omega_{\rm m}=0.3$, corresponding to a physical scale of $7.62~{\rm kpc~arcsec^{-1}}$ ($0\farcs03 \approx 230$~pc) at $z=3.111$.

\section{Observations and Data Reduction}
\label{sec:data}

\subsection{ALMA observations}
\label{sec:alma}

W2305$-$0039 is a \emph{WISE}-selected Hot DOG with a spectroscopic redshift $z=3.111$ \citep{Tsai2015,Martin2024}.
Infrared SED decomposition yields an AGN luminosity of $\log(L_{\rm AGN}/\Lsun)=14.14\pm0.02$ \citep{Sun2024}. 
While the statistical uncertainty on this value is small, the error budget is dominated by a fractional systematic error of $\sigma_{\rm sys}=0.33$ within the BayeSED framework \citep{Sun2024,2014ApJS..215....2H}.
We therefore adopt a total uncertainty of $\pm0.1$~dex in $\log L_{\rm AGN}$, including the systematic error.

\begin{deluxetable*}{llrcl}
\tablecaption{Summary of ALMA observations \label{tab:alma}}
\tablehead{\colhead{Line} & \colhead{Receiver} & \colhead{Integration} & \colhead{Max.\ angular scale} & \colhead{Project ID}}
\startdata
CO(7--6), \ci, 1.4\,mm cont. & Band 5 & 265 min & 0\farcs64 & 2024.1.01175.S (PI: K.~Tadaki) \\
CO(7--6), \ci, 1.4\,mm cont. & Band 5 & 5 min & 13$"$ & 2022.1.00353.S (PI: A.~Manuel) \\
CO(7--6), \ci, 1.4\,mm cont. & Band 5 & 5 min & 4\farcs5 & 2021.1.00168.S (PI: F.~Timothy) \\
CO(11--10), 1.0\,mm cont. & Band 7 & 280 min & 0\farcs48 & 2024.1.01175.S (PI: K.~Tadaki) \\
CO(11--10), 1.0\,mm cont. & Band 7 & 7 min & 4\farcs2 & 2021.1.00168.S (PI: F.~Timothy) \\
\enddata
\end{deluxetable*}

We observed W2305$-$0039 with ALMA, targeting the CO $J$=7--6 and CO $J$=11--10 transitions, both of which were detected at $\sim$1\arcsec\ resolution in Paper~I \citep{2026arXiv260223521T}. 
The CO(7--6) and \ciitwo\ (hereafter [\ion{C}{1}]) lines were covered simultaneously in Band~5, while CO(11--10) was observed in Band~7. 
Our analysis uses multiple ALMA projects to cover a wide range of spatial scales (Table~\ref{tab:alma}). 
All imaging products and the dynamical modeling in this paper use only the high-resolution data from project 2024.1.01175.S (PI: K. Tadaki). 
The lower-resolution archival datasets from 2021.1.00168.S and 2022.1.00353.S are used solely in the visibility-domain size fits of Section \ref{sec:methods_uv}, where they help constrain the extended flux at short $uv$ distances.

For the high-resolution project (2024.1.01175.S), the spectral-line setup employed two basebands configured in frequency-division mode (FDM) using 4-bit correlator mode.
To reduce data volume, the visibilities were spectrally averaged to a channel width of 7.8~MHz, corresponding to velocity resolutions of 12~\kms\ for the CO(7--6)+[\ion{C}{1}] observations and 8~\kms\ for the CO(11--10) observations. 
These velocity resolutions are much smaller than the line widths set by the rotation velocity and intrinsic velocity dispersion of W2305$-$0039 and therefore do not limit our kinematic analysis (Section~\ref{sec:kinematics}). 
The remaining two basebands were used for continuum measurements in FDM 2-bit mode, providing a total effective continuum bandwidth of 3.75~GHz ($1.875~{\rm GHz} \times 2$), with representative continuum center frequencies of 208.951~GHz (1.4~mm) in Band~5 and 296.642~GHz (1.0~mm) in Band~7. 

For the project 2024.1.01175.S, we re-ran the Common Astronomy Software Application (\texttt{CASA}) v6.6.6 pipeline \citep{CASA2022} on the raw visibility data rather than using the archive-provided pipeline-calibrated products. 
This re-processing was needed to work around a known issue in \texttt{hifa\_flagdata}, where FDM spectral windows with a small number of channels can be misidentified as TDM windows and have their edge channels flagged unnecessarily. 
We used a modified procedure that skipped the \texttt{edgespw} step for the affected spectral windows.
Additional datasets from projects 2021.1.00168.S and 2022.1.00353.S were taken from the ALMA Science Archive and used as pipeline-calibrated visibility data. 

\begin{figure*}
\centering
\figfile[scale=1.0]{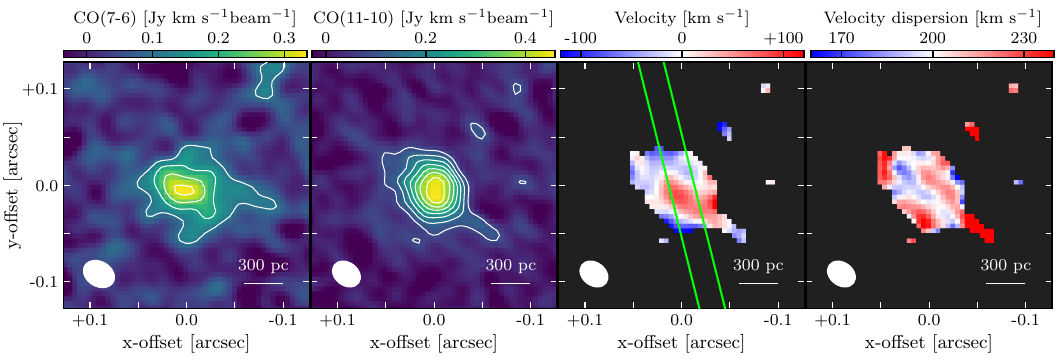}
\caption{Morphology and kinematics of W2305$-$0039 at $z=3.111$. ALMA maps of \co(7--6) integrated intensity, \co(11--10) integrated intensity, velocity field of the \co(11--10) line, and velocity dispersion. 
Contours are plotted every $2\sigma$, starting at $3\sigma$. 
The filled circle indicates the synthesized beam. 
The green lines indicate the orientation of the slit used to extract the PV diagram shown in Figure~\ref{fig:pv}.}
\label{fig:moments}
\end{figure*}

\subsection{Imaging}
\label{sec:imaging}

We imaged only the high-resolution dataset from project 2024.1.01175.S.
For the line cubes, continuum subtraction was performed in the visibility domain using the \texttt{CASA/uvcontsub\_old} task, fitting a first-order polynomial to line-free channels across all four basebands to estimate and remove the continuum.
The line cubes were then imaged with the \texttt{CASA/tclean} task using a channel width of 100~\kms\ and a spatial pixel scale of 0\farcs005.
We first generated initial images using Briggs weighting with robust~$=2.0$ and a $uv$-taper of 0\farcs05, and constructed masks automatically with \texttt{auto-multithresh} \citep{Kepley2020}. 
Using these masks, we then produced the final images with Briggs weighting robust~$=0.5$ without $uv$-taper. 
Deconvolution employed multiscale CLEAN \citep{Cornwell2008}, and cleaning was stopped at a threshold of $\approx$1.5 times the measured rms noise. 
The resulting synthesized beam for the CO(7--6) cube is $0\farcs0352 \times 0\farcs0269$ (P.A.~$= 59.3^{\circ}$) with an rms sensitivity of 63.0~$\mu$Jy~beam$^{-1}$.
For the CO(11--10) cube, the beam is $0\farcs0325 \times 0\farcs0250$ (P.A.~$= 53.5^{\circ}$) with an rms of 61.3~$\mu$Jy~beam$^{-1}$.
Moment-0 (velocity-integrated intensity), moment-1 (velocity field), and moment-2 (velocity dispersion) maps were produced using the \texttt{CASA/immoments} task, integrating over the velocity range from $-450$ to $+450$~\kms\ (Figure~\ref{fig:moments}). For the moment-1 and moment-2 maps, we masked pixels where the corresponding moment-0 intensity is below $3\sigma$.

\begin{figure}
\centering
\figfile[scale=1.0]{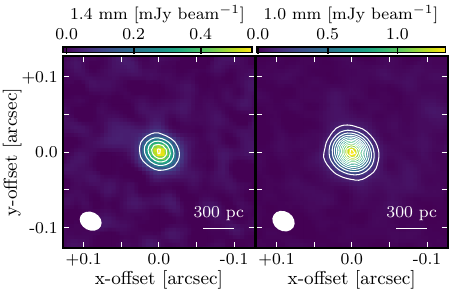}
\caption{Dust continuum emission in W2305$-$0039 at observed-frame 1.4~mm (Band~5; left) and 1.0~mm (Band~7; right).
Contours are shown at 10, 20, 30, \ldots, 80$\sigma$ (left) and 10, 20, 30, \ldots, 120$\sigma$ (right).}
\label{fig:cont}
\end{figure}

Continuum images were constructed from the two line-free basebands in each band. 
For both Band~5 (1.4~mm) and Band~7 (1.0~mm), the continuum data were imaged using the same strategy as for the line cubes, including the same spatial pixel scale, weighting schemes, and CLEAN masking procedure. 
The resulting 1.4~mm continuum image has a synthesized beam of $0\farcs0303 \times 0\farcs0241$ (P.A.~$= 63.0^{\circ}$) and an rms sensitivity of 10.8~$\mu$Jy~beam$^{-1}$, while the 1.0~mm continuum image has a beam of $0\farcs0311 \times 0\farcs0258$ (P.A.~$= 57.6^{\circ}$) and an rms of 10.9~$\mu$Jy~beam$^{-1}$. 
Both continuum maps are shown in Figure~\ref{fig:cont}.

\FloatBarrier

\section{Resolved structure and excitation of the molecular gas}
\label{sec:methods}

\subsection{Size measurements}
\label{sec:methods_uv}

To quantify intrinsic source sizes and radial excitation without beam-smearing biases, we modeled the spatial distributions of the \co(7--6), \co(11--10), and [\ion{C}{1}] line emission, as well as the dust continuum, directly in the visibility domain with \texttt{UVMULTIFIT} \citep{MartiVidal2014}.
We adopted a two-component model consisting of two exponential disks: a compact, elliptical disk and an extended, circular disk. 
The two components were constrained to share a common center, while the compact component was allowed to have a free axis ratio and position angle. 
For the line datasets, we first averaged the calibrated visibilities over the velocity range from $-450$ to $+450$~\kms\ using the \texttt{CASA/mstransform} task, producing integrated line visibility data for model fitting. 
We fitted the CO(11--10) visibilities, which have the highest signal-to-noise ratio among the three line datasets, to constrain the compact-disk geometry. 
We then fixed the compact-disk axis ratio and position angle to these CO(11--10) best-fit values when fitting the CO(7--6) and [\ion{C}{1}] visibilities, to stabilize the decomposition between the compact and extended components. 
For the dust continuum datasets, we fitted the visibilities with the same two-disk model while keeping the compact-component axis ratio and position angle as free parameters.

\begin{figure*}
\centering
\figfile[scale=1.0]{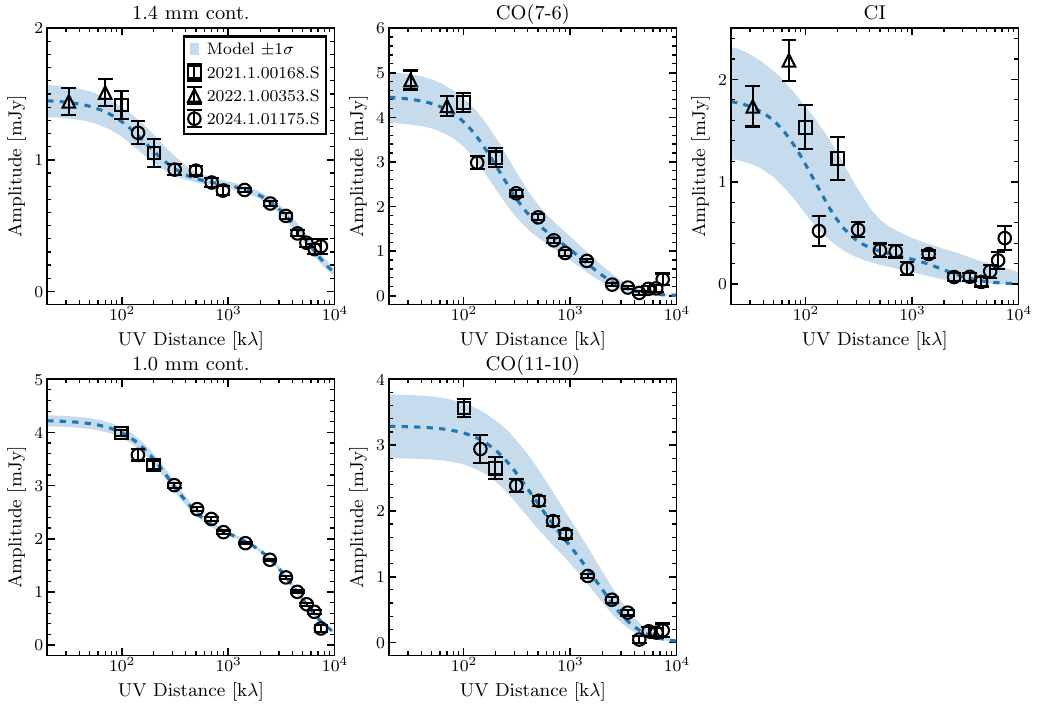}
\caption{Azimuthally averaged visibility amplitudes as a function of $uv$ distance for the continuum and line datasets. 
Data points show the binned amplitudes (error bars, $1\sigma$) for the individual ALMA projects, as indicated by the symbols. 
Dashed curves show the best-fitting two-component exponential-disk models, circularized for the azimuthal averaging. 
Shaded regions indicate the $1\sigma$ model uncertainty propagated from the fitted parameter uncertainties via Monte Carlo sampling.
}
\label{fig:visamp}
\end{figure*}

A comparison between the observed visibility amplitudes and the best-fitting models is shown in Figure~\ref{fig:visamp}.
The best-fitting parameters for all line and continuum datasets are summarized in Table~\ref{tab:uvfit}.
The resulting circularized effective radii of the compact-disk components show that the higher-excitation CO emission is more compact than the lower-excitation emission, with $R_{\rm e} = 173 \pm 30$~pc for CO(11-10) and $R_{\rm e} = 259 \pm 58$~pc for CO(7-6). 
The two sizes are still consistent at about the 1.3$\sigma$ level, but the same trend is seen in the visibility profiles of Figure \ref{fig:visamp}, where the CO(11-10) amplitude drops off more slowly with $uv$ distance than CO(7-6).
The 1.0~mm dust continuum emission is even more centrally concentrated, with $R_{\rm e} = 77 \pm 3$~pc, indicating that the dominant energy source is confined to the innermost region and is likely associated with an AGN.

To visualize departures from the axisymmetric two-disk model, we subtracted the best-fitting visibility model from the CO(7--6) visibilities and imaged the residuals using \texttt{CASA/tclean} with \texttt{niter}=0 (i.e., a dirty residual map). 
For this imaging, we adopted Briggs weighting with \texttt{robust}~$=2.0$ and a $uv$-taper of $0\farcs05$ to highlight extended low-level emission.
The residual map has an rms of 50.1~$\mu$Jy~beam$^{-1}$ and a synthesized beam of $0\farcs098 \times 0\farcs081$ (P.A.~$= 84.7^{\circ}$). 
We use this residual map to assess low-level non-axisymmetric structure that is not captured by the two-component parametrization (see Section~\ref{sec:discussion}).

\begin{deluxetable*}{llcccc}
\tablecaption{Visibility-plane two-component exponential-disk fits \label{tab:uvfit}}
\tablehead{\colhead{Dataset} & \colhead{Component} & \colhead{Flux (mJy)} & \colhead{$R_{\rm e}$ (arcsec)} & \colhead{$q$} & \colhead{P.A. (deg)}}
\startdata
CO(11--10) & Extended & $1.59\pm0.36$ & $0.117\pm0.044$ & 1.0 (fixed) & 0 (fixed) \\
CO(11--10) & Compact  & $1.70\pm0.34$ & $0.026\pm0.005$ & $0.76\pm0.15$ & $14\pm19$ \\
CO(7--6)   & Extended & $3.01\pm0.49$ & $0.214\pm0.057$ & 1.0 (fixed) & 0 (fixed) \\
CO(7--6)   & Compact  & $1.46\pm0.33$ & $0.035\pm0.007$ & 0.76 (fixed) & 14 (fixed) \\
\ci    & Extended & $1.52\pm0.55$ & $0.373\pm0.216$ & 1.0 (fixed) & 0 (fixed) \\
\ci    & Compact  & $0.31\pm0.16$ & $0.026\pm0.018$ & 0.76 (fixed) & 14 (fixed) \\
1.4\,mm cont. & Extended & $0.63\pm0.12$ & $0.263\pm0.074$ & 1.0 (fixed) & 0 (fixed) \\
1.4\,mm cont. & Compact  & $0.83\pm0.03$ & $0.008\pm0.001$ & $1.0\pm0.12$ & $44\pm1$ \\
1.0\,mm cont. & Extended & $2.10\pm0.10$ & $0.158\pm0.011$ & 1.0 (fixed) & 0 (fixed) \\
1.0\,mm cont. & Compact  & $2.13\pm0.03$ & $0.010\pm0.001$ & $0.85\pm0.05$ & $3\pm7$ \\
\enddata
\end{deluxetable*}

\subsection{Radial CO excitation profile}\label{sec:radial_profile}

\begin{figure}
\centering
\figfile[scale=1.0]{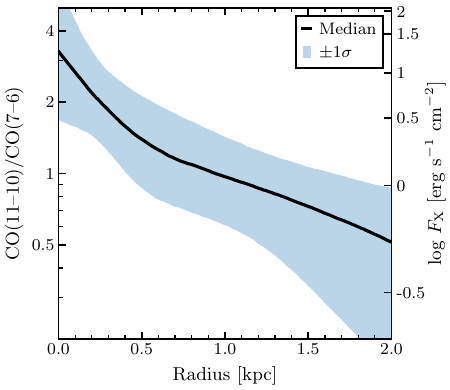}
\caption{Radial profile of the \co(11--10)/\co(7--6) luminosity ratio derived from the visibility modeling. 
The shaded region indicates the $1\sigma$ uncertainty from Monte Carlo error propagation.
The left axis shows the line ratio, while the right axis gives the corresponding incident X-ray flux predicted by the XDR model ($M_{\rm GMC} = 10^{4.5}~\Msun$; \citealt{Esposito2024}). 
The systematic uncertainty in $F_{\rm X}$ from the choice of GMC mass is about $\pm$0.1 dex (see Section \ref{sec:methods_xdr}).
}
\label{fig:ratio}
\end{figure}

Based on the best-fit CO(11--10) and CO(7--6) models, we derived a radial profile of the CO(11--10)-to-CO(7--6) luminosity ratio (Figure~\ref{fig:ratio}).
For each transition, we reconstructed the intrinsic surface-brightness distribution using the two-component disk model obtained from the visibility analysis. 
The total surface-brightness profile for each line was obtained as the sum of the compact and extended components assuming circularized radii. 
To propagate uncertainties from the visibility fitting into the radial profiles, we performed a Monte Carlo error analysis: the best-fit parameters of each transition were drawn from Gaussian distributions with their $1\sigma$ uncertainties, generating 5,000 realizations of the surface-brightness profile. 
For each realization, the CO(11--10)/CO(7--6) luminosity ratio profile was computed, and we adopted the median and the 16th--84th percentile range at each radius.
The ratio rises steeply above unity within the central $\sim$500~pc and declines at larger radii. 

\subsection{Interpretation with PDR/XDR models}\label{sec:methods_xdr}

To interpret the origin of this extreme excitation, we compared the observed ratio with predictions from PDR models powered by star formation and XDR models associated with an AGN. 
We used \texttt{galaxySLED} \citep{Vallini2019,Esposito2024}, which builds on a semi-analytical giant molecular cloud (GMC) framework and couples radiative transfer calculations of PDR and XDR to a physically motivated description of GMC substructure. 
Rather than treating the molecular gas as a single-density slab, each GMC is represented as a collection of clumps drawn from a turbulence-regulated density probability distribution, so that dense gas capable of powering high-$J$ emission is included without imposing an ad hoc constant clump density. 
For each clump, CO line emissivities are computed using \texttt{CLOUDY} \citep{Ferland2017} and then integrated over the internal clump density distribution within a single GMC to obtain the total CO luminosity in each transition. 
We adopted a median GMC mass of $M_{\rm GMC} = 10^{4.5}~\Msun$ with a mean number density of 550~cm$^{-3}$ \citep{Esposito2024} and generated representative CO spectral line energy distributions (SLEDs) for an extreme PDR field ($\log G_{0} = 6$) and for XDR models spanning $\log(F_{\rm X}/{\rm erg~s^{-1}~cm^{-2}}) = -1$ to $2$. 

Even PDR models with exceptionally intense ultraviolet radiation fields ($G_{0}=10^{6}$) predict CO(11--10)/CO(7--6)~$\simeq 0.02$, far below the observed values, implying an excitation state that cannot be explained by star formation alone \citep{Werf2010,Gallerani2014,Tadaki2025}. 
In contrast, XDR models reproduce the measured excitation in the central region, provided that the molecular gas is exposed to sufficiently strong X-ray irradiation ($F_{\rm X} \gtrsim 1~{\rm erg~s^{-1}~cm^{-2}}$).
This conclusion is robust against the choice of GMC model. 
We repeated the calculation with GMC models of different masses ($M_{\rm GMC} = 10^{3.1}~\Msun$ and $10^{5.9}~\Msun$), and in all three cases the PDR-predicted ratio remains well below unity even at $\log G_{0} = 6$, while only XDR models can reach the observed values of CO(11--10)/CO(7--6)$>$1.
For an observed CO(11--10)/CO(7--6) ratio above unity, the incident X-ray flux required by the XDR model varies by about $\pm$0.1 dex across the three GMC models, which does not affect our interpretation of AGN-driven X-ray heating.

At a radius of $\sim$1~kpc, converting the threshold $F_{\rm X} \gtrsim 1~{\rm erg~s^{-1}~cm^{-2}}$ to a luminosity gives a minimum of $\log(L_{\rm X}/{\rm erg~s^{-1}}) \simeq 44.1$. 
Given the AGN luminosity $\log(L_{\rm AGN}/{\rm erg~s^{-1}}) = 47.7$ (Section~\ref{sec:alma}), this implies $\log(L_{\rm X}/L_{\rm AGN}) \simeq -3.6$.
This ratio represents a lower limit on the intrinsic X-ray-to-bolometric fraction, because any attenuation within the compact, obscured circumnuclear medium would reduce the flux reaching the molecular disk. 
Even for a moderate gas column density of $N_{\rm H} \sim 10^{23}~{\rm cm^{-2}}$, the intrinsic X-ray luminosity could be higher by about an order of magnitude \citep{Maloney1996}, raising $\log(L_{\rm X}/L_{\rm AGN})$ to $\sim$$-2.5$ and approaching values reported for luminous obscured quasars \citep{Vito2018}. 
The spatial coincidence of the high-excitation region with the compact dust continuum emission further supports this interpretation, pointing to heating by a deeply obscured AGN rather than a galaxy-wide starburst.

\subsection{Molecular gas mass estimates}
\label{sec:methods_gasmass}

We estimated the molecular gas mass from the \ci\ emission line following standard procedures that treat atomic carbon as a tracer of the bulk molecular interstellar medium \citep{Weiss2005}, using the \ci\ fluxes of the compact and extended components obtained from the visibility-plane decomposition (Table~\ref{tab:uvfit}).
We assumed optically thin emission and local thermodynamic equilibrium (LTE), and inferred the total mass of neutral atomic carbon from the \ci\ luminosity assuming a single excitation temperature and the appropriate partition function.
The atomic carbon mass was then converted into a molecular gas mass by adopting a carbon abundance relative to molecular hydrogen and applying a standard correction factor of 1.36 for helium \citep{2004MNRAS.351..147P,2017MNRAS.466.2825B}. 
For the compact component, we adopted a fiducial excitation temperature of $T_{\rm ex}=100$~K to reflect its warm environment and a neutral carbon abundance of $8\times10^{-5}$ relative to H$_2$, consistent with observationally inferred values in high-redshift dusty galaxies \citep{Walter2011}. 
Such a high abundance is plausibly associated with AGN-irradiated environments, where X-ray--driven chemistry can partially dissociate CO and shift the carbon phase balance toward \ci\ \citep{MeijerinkSpaans2005}. 
Under these assumptions, we derived a molecular gas mass of $M_{\rm gas} =(6.3 \pm 3.5)\times 10^{9}~\Msun$ for the compact disk. 
For a direct comparison, we applied the same assumptions to the extended disk, yielding a molecular gas mass of $M_{\rm gas} =(3.1\pm1.1)\times 10^{10}~\Msun$. 
We note, however, that the extended disk is likely characterized by cooler gas and a lower carbon abundance closer to commonly adopted Galactic values ($\sim$$2\times10^{-5}$; \citealt{Frerking1989}).
Adopting such conditions would increase the inferred gas mass, so this estimate should be regarded as a lower limit.

\begin{figure*}
\centering
\figfile[scale=1.0]{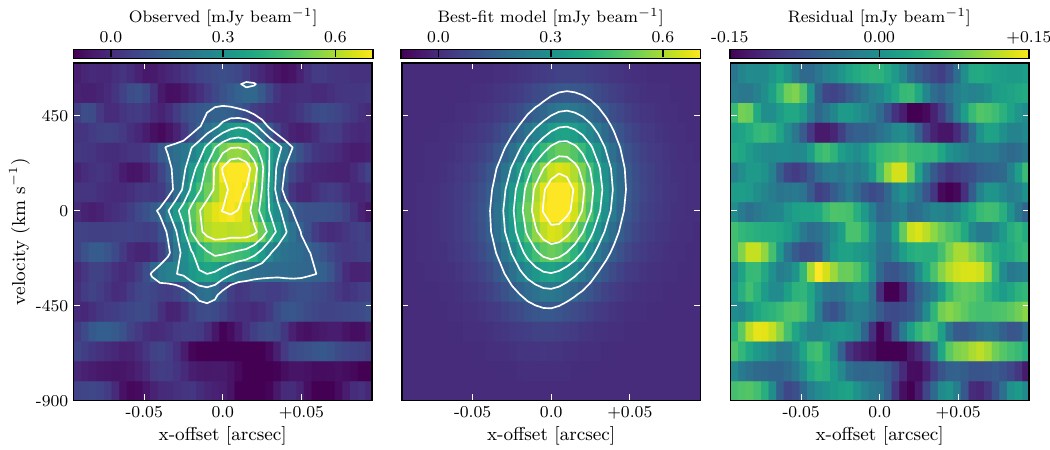}
\caption{Position--velocity modeling of the nuclear \co(11--10) emission. Panels show the observed, best-fitting model, and residual PV diagrams extracted along the kinematic major axis. Contours indicate signal-to-noise levels of $3$--$13\sigma$ in steps of $2\sigma$.}
\label{fig:pv}
\end{figure*}

\section{Nuclear kinematics}
\label{sec:kinematics}

\subsection{PV extraction}
\label{sec:methods_pv}

Motivated by the compact CO(11--10) distribution and its extreme excitation, we investigated the kinematics of the molecular gas in the nuclear region of W2305$-$0039.
The velocity field and velocity dispersion maps (Figure~\ref{fig:moments}) reveal a highly turbulent state in which random motions dominate over ordered rotation within the central $\simeq$300~pc. 
Such dispersion-dominated kinematics are similar to those observed in other luminous dust-obscured galaxies at $z \simeq 3$--$4$ \citep{Martin2024,DiazSantos2016,Liao2025}. 
Even at our 200~pc resolution, the velocity dispersion remains elevated across multiple synthesized beams, ruling out beam smearing as the primary cause and indicating intrinsically non-circular motions.
Although the kinematics are dominated by large intrinsic velocity dispersion, we detected a weak velocity gradient along the major axis of the compact CO(11--10) disk, enabling the black-hole mass to be constrained through forward modeling that accounts for both rotation and dispersion. 
To isolate this ordered component, we extracted a position--velocity (PV) diagram along the morphological major axis (Figure~\ref{fig:pv}).
The PV slice was defined using the position angle of the compact disk and a slit width of three spatial pixels.
The resulting PV diagram is strongly centrally concentrated and remains broad in velocity at all offsets, consistent with dispersion-dominated nuclear gas motions.
A marginal velocity gradient is nonetheless present across the central region, providing the ordered signal we model to constrain the enclosed dynamical mass.

\subsection{Dynamical forward modeling}
\label{sec:methods_dyn}

To constrain the gravitational potential, we forward-modeled the observed PV diagram with \texttt{KinMS} \citep{Davis2013}, a publicly available code that generates synthetic three-dimensional data cubes of arbitrary rotating gas disks. 
Given an input radial surface-brightness profile, an intrinsic velocity dispersion, and a circular velocity curve, \texttt{KinMS} produces a Monte Carlo realization of the gas distribution and then convolves it with the observing beam and channel width, yielding a model cube that can be compared directly with the data.
The model assumed an axisymmetric rotating disk with (i) an exponential radial surface-brightness distribution and (ii) a spatially uniform intrinsic velocity dispersion. 
We computed the circular velocity curve as the sum of a central point mass (black hole) and an exponential disk mass component. 
To evaluate the disk contribution efficiently at each MCMC step, we approximated the exponential disk surface-density profile with a multi-Gaussian expansion (MGE), i.e., a sum of concentric coaxial Gaussian components \citep{Cappellari2008}.
This MGE representation allows the disk circular velocity curve to be computed rapidly using the axisymmetric MGE formalism, which we then combined with the central point-mass term to obtain the total circular velocity curve. 
For the disk component, both the total mass $M_{\rm disk}$ and effective radius $R_{\rm e,disk}$ were treated as free parameters. 
We note that the exponential disk used here differs in purpose from the two-component exponential-disk decomposition of Section \ref{sec:methods_uv}. 
In Section \ref{sec:methods_uv}, the two-disk model is an empirical description of the spatial surface-brightness distribution of each tracer, used only to measure sizes and to separate compact and extended flux. 
In contrast, the disk in this section represents the enclosed mass distribution that contributes to the circular velocity together with the central black hole.
We further keep the CO(11-10) surface-brightness profile (the input intensity distribution for \texttt{KinMS}) separate from this disk mass profile, because CO(11-10) traces warm, highly excited gas that does not necessarily follow the bulk mass distribution.
For each trial parameter set, \texttt{KinMS} generated a beam-convolved cube on a fine spatial grid, from which we extracted a model PV diagram using the same slit geometry as applied to the data.

\begin{figure}
\centering
\figfile[scale=1.0]{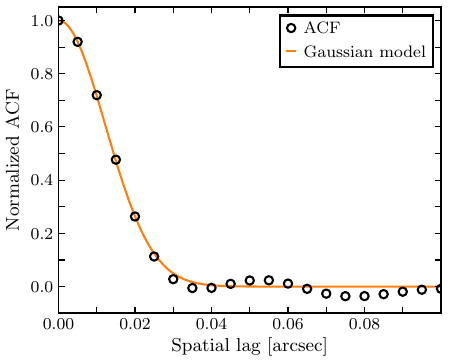}
\caption{Spatial noise correlations in the PV diagram estimated from an emission-free noise slice.}
\label{fig:acf}
\end{figure}

We compared the observed and model PV diagrams using a generalized least-squares (GLS) likelihood that accounts for correlated noise along the spatial axis. We defined a fixed detection mask in the PV plane using a signal-to-noise threshold of ${\rm S/N} > 3$ and evaluated the likelihood only within this mask. 
To characterize the noise in the PV diagram, we extracted an additional PV slice from an emission-free region of the same cube using an identical slit geometry.
From this noise PV we obtained a robust rms estimate $\sigma_{\rm rms}$ and quantified the correlation along the spatial axis. 
Specifically, we computed the one-dimensional autocorrelation function (ACF) of the noise PV as a function of spatial lag, averaged over velocity channels, and fitted the ACF with a Gaussian (Figure~\ref{fig:acf}) to estimate an effective correlation length $\sigma_{x}$. 
This yielded $\sigma_{x} = 0\farcs0123$ (FWHM~$= 0\farcs0288$), consistent with the synthesized beam size projected onto the slit direction. 
Because the 0\farcs005 pixel sampling oversamples the synthesized beam by a factor of 5--6 across the beam FWHM, adjacent pixels are strongly correlated \citep{Tsukui2023}.
In contrast, we assumed that different velocity channels are statistically independent, as the PV modeling was performed on a cube binned to 100~\kms\ channels, much coarser than the native spectral resolution ($\sim$8~\kms).

We therefore modeled the spatial covariance between pixels $i$ and $j$ in each velocity channel as a Gaussian correlation kernel plus a small white-noise term,
\begin{equation}
C_{ij} = \sigma_{\rm rms}^{2}\left[ (1 - \eta)\exp\!\left( - \frac{(x_{i} - x_{j})^{2}}{2\sigma_{x}^{2}} \right) + \eta\,\delta_{ij} \right],
\end{equation}
where $\eta$ controls the fractional contribution of an uncorrelated component. 
Although the noise ACF indicates that the spatial correlations are well described by a single Gaussian kernel with negligible white-noise contribution, we adopted a small non-zero floor ($\eta = 0.01$) purely as a numerical regularization term to stabilize the covariance inversion and prevent ill-conditioning in the presence of strong spatial correlations.
Because the absolute flux normalization of the forward-model PV is degenerate with the normalization of the assumed surface-brightness profile, we allowed an overall multiplicative scaling between the model and data and determined it analytically for each trial model within the GLS framework. 
This procedure emphasizes the shape of the PV diagram (spatial extent, velocity gradient, and line broadening) rather than the absolute flux scale.

\begin{figure*}
\centering
\figfile[scale=1.0]{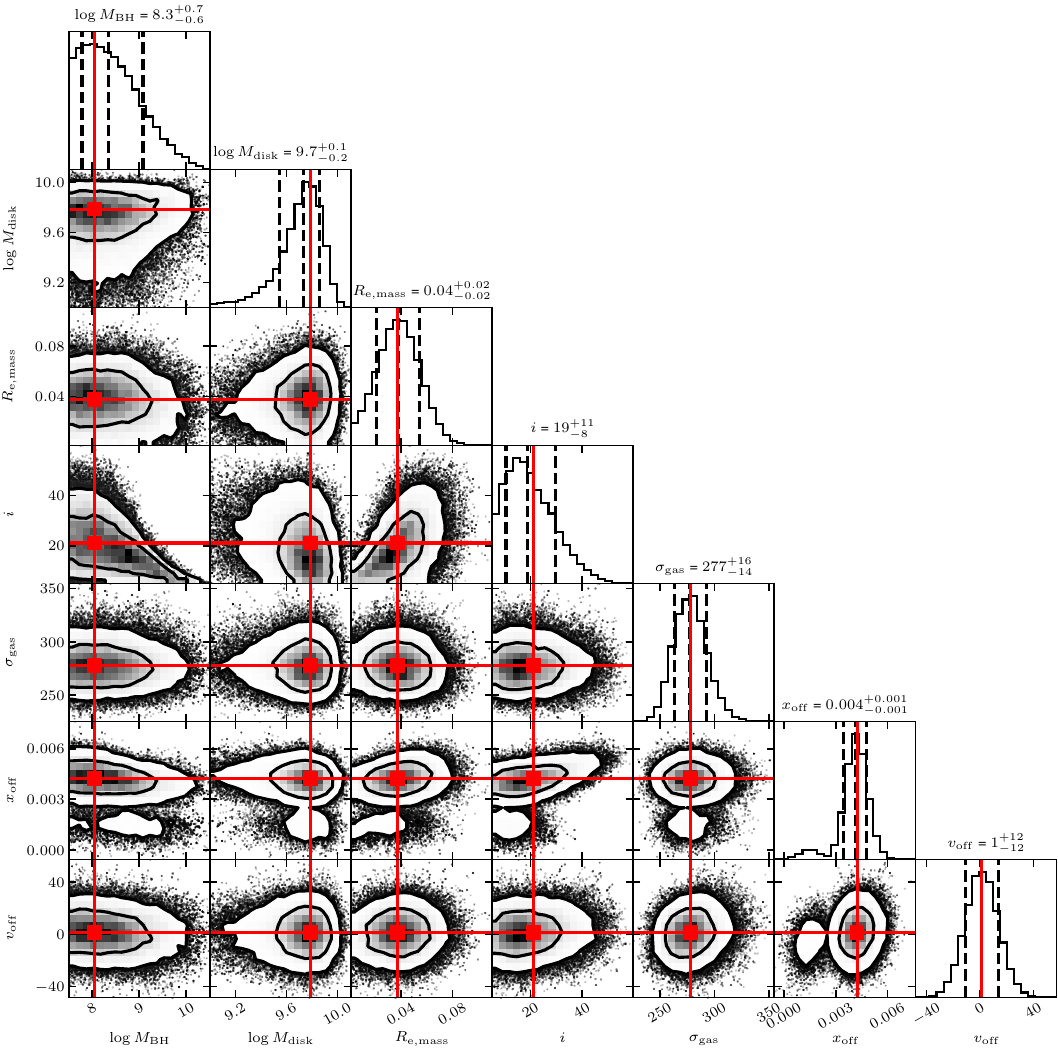}
\caption{
Posterior distributions from the MCMC fitting of the PV model. 
The one-dimensional panels report the median values and the 16th--84th percentile credible intervals. 
The red vertical and horizontal lines indicate the best-fitting parameter values corresponding to the maximum posterior probability.
}
\label{fig:corner}
\end{figure*}

We explored the posterior distribution of the dynamical parameters using the Markov chain Monte Carlo sampler \texttt{emcee} \citep{ForemanMackey2013}. 
The sampled parameters are the black-hole mass $M_{\rm BH}$, the disk mass $M_{\rm disk}$, the effective radius of the disk mass component $R_{\rm e,disk}$, the disk inclination $i$, the intrinsic velocity dispersion $\sigma_{\rm gas}$, and spatial and velocity offsets. 
We adopted broad uniform bounds for all sampled parameters and incorporated external information via informative priors on $i$, $M_{\rm disk}$, and $R_{\rm e,disk}$ where appropriate. For the inclination, we used the deconvolved apparent minor-to-major axial ratio $q_{\rm obs}$ of the compact CO(11--10) disk measured from the visibility modeling. 
For an axisymmetric disk with intrinsic thickness $q_{\rm int} = 0.3$ viewed at inclination $i$ (with $i = 0^{\circ}$ face-on and $i = 90^{\circ}$ edge-on), the expected projected axial ratio is $q_{\rm model}(i) = \sqrt{q_{\rm int}^{2} + (1 - q_{\rm int}^{2})\cos^{2}i}$.
We applied a Gaussian prior on $q_{\rm model}(i)$ centered on $q_{\rm obs}$ with the measurement uncertainty and multiplied it by the geometric prior $p(i) \propto \sin i$ corresponding to an isotropic distribution of disk orientations. We further applied Gaussian priors on $M_{\rm disk}$ and $R_{\rm e,disk}$, centered on the \ci-inferred gas mass and the \ci-based size constraint, respectively. These \ci-motivated priors are intended to anchor the mass component to a tracer of the bulk molecular reservoir, whereas CO(11--10) primarily traces a warmer, highly excited phase. 
The posterior distributions are shown in Figure~\ref{fig:corner}.

The PV modeling yields a dynamical black-hole mass of $\log(M_{\rm BH}/\Msun) = 8.3^{+0.7}_{-0.6}$
and a large intrinsic gas velocity dispersion of $\sigma_{\rm gas} = 277^{+16}_{-14}$~\kms. 
This dynamical estimate is $\sim$0.5--1.0~dex lower than the single-epoch virial black-hole masses typically inferred from rest-frame ultraviolet \ion{Mg}{2} or \ion{C}{4} broad lines in other luminous dust-obscured galaxies at $z \simeq 3$--$4$ \citep{Tsai2018,Luo2025}. 
UV single-epoch virial estimates can be biased by non-virial broad-line components, particularly for \ion{C}{4} in outflow-dominated systems \citep{Coatman2017}, and are also sensitive to extinction corrections in obscured AGN \citep{Ricci2022}. 
An independent dynamical measurement therefore provides a valuable benchmark for black-hole masses in these systems.

The MCMC posterior distributions show a pronounced degeneracy between the black-hole mass and disk inclination. 
The PV diagram alone does not tightly constrain $i$, and the inferred black-hole mass is sensitive to the inclination prior. 
To assess how strongly our $M_{\rm BH}$ constraint depends on the geometric orientation prior, we performed an additional fitting in which we removed the isotropic-orientation weighting $p(i) \propto \sin i$ from the prior.
Removing the $\sin i$ factor is not physically motivated for a randomly oriented population, because it artificially increases the prior probability of nearly face-on configurations, which are intrinsically rare. We therefore do not adopt this prior choice as our fiducial model, but instead use it as a deliberate stress test designed to maximize the allowed contribution of low-inclination solutions and thereby obtain a conservative upper envelope on $M_{\rm BH}$. 
In this test, the posterior shifts to smaller inclinations ($i = 14^{+11}_{-6}$~deg) and correspondingly allows a higher black-hole mass with $\log(M_{\rm BH}/\Msun) = 8.5^{+0.8}_{-0.7}$.
The 84th percentile corresponds to $\log(M_{\rm BH}/\Msun) \approx 9.4$, which we treat as a conservative upper bound under an intentionally permissive inclination prior. 

\section{Discussion}
\label{sec:discussion}

Combining our dynamical constraint on the black-hole mass with the independently estimated AGN luminosity of $\log(L_{\rm AGN}/{\rm erg~s^{-1}}) = 47.7 \pm 0.1$ \citep{Sun2024} provides a direct test of the accretion state of W2305$-$0039.
For the dynamically inferred black-hole mass, the corresponding Eddington luminosity, $\log(L_{\rm Edd}/{\rm erg~s^{-1}}) = 46.4^{+0.7}_{-0.6}$, implies an Eddington ratio well above unity ($\log\lambda_{\rm Edd} = 1.3^{+0.6}_{-0.7}$, i.e., $\lambda_{\rm Edd} \gtrsim 4$ even at the $1\sigma$ lower bound). 
Moreover, even adopting the conservative upper bound of $\log(M_{\rm BH}/\Msun) \approx 9.4$ obtained from the stress test without the $\sin i$ prior (Section~\ref{sec:methods_dyn}), the Eddington ratio remains above unity, demonstrating that our conclusion of super-Eddington accretion is robust against the choice of inclination prior.
Accretion above the classical Eddington limit has long been proposed as a mechanism for the rapid assembly of supermassive black holes in the early Universe, and Eddington ratios at or above unity have been suggested for luminous dust-obscured galaxies and quasars based primarily on ultraviolet single-epoch virial mass estimates \citep{Tsai2018,Banados2021}. 
By anchoring the black-hole mass with molecular-gas dynamics, our measurement provides a more direct assessment of whether W2305$-$0039 is accreting above the classical Eddington limit.

This super-Eddington state implies rapid black-hole growth but cannot be sustained indefinitely at fixed AGN luminosity, because the Eddington ratio decreases as the black hole gains mass. 
For a thin-disk estimate with $\epsilon=0.1$ \citep{Marconi2004}, the accretion rate is $\dot{M}_{\rm acc}=L_{\rm AGN}/(\epsilon\,c^{2})\simeq 90~\Msun~{\rm yr}^{-1}$ at the nominal luminosity.
Retaining the fiducial $\epsilon=0.1$, the Eddington ratio would drop to unity once the black-hole mass grows by a factor of $\sim$20 to $M_{\rm BH} \approx 4 \times 10^{9}~\Msun$, which at the present accretion rate would take $\lesssim$50~Myr.
We therefore assessed whether the circumnuclear gas reservoir can sustain the present luminosity over this timescale. 
The compact \ci\ disk with $R_{\rm e} = 173 \pm 120$~pc contains $M_{\rm gas} = (6.3 \pm 3.5) \times 10^{9}~\Msun$ of molecular gas (Section~\ref{sec:methods_gasmass}), implying a nominal gas depletion time of $\sim$70~Myr at the current accretion rate.
The compact reservoir can therefore in principle fuel substantial black-hole growth in a single episode. 
However, if a non-negligible fraction of the compact-disk gas is consumed by nuclear star formation or expelled in outflows, sustaining the present hyperluminous output for tens of Myr would likely require additional gas inflow from larger radii.

\begin{figure}
\centering
\figfile[scale=1.0]{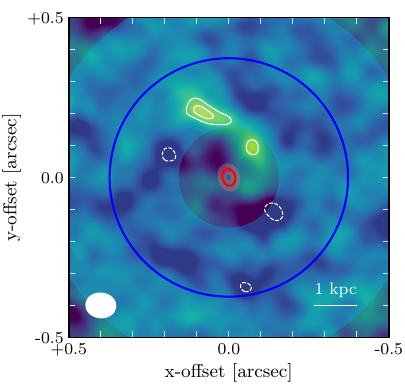}
\caption{Residual \co(7--6) emission after subtracting the best-fit two-component visibility model, illustrating low-level non-axisymmetric structure. Contours are shown at $\pm3\sigma$ and $\pm4\sigma$, with negative contours dashed.
The blue and red ellipses indicate the best-fit extended and compact disk components traced by \ci, respectively. 
The shaded regions represent their 1$\sigma$ uncertainties in \ci~size.
}
\label{fig:residual}
\end{figure}

Beyond the compact disk, W2305$-$0039 also hosts a more extended \ci\ disk with $R_{\rm e} = 2.8 \pm 1.6$~kpc, containing at least $M_{\rm gas} =(3.1 \pm 1.1) \times 10^{10}~\Msun$ of molecular gas (Section~\ref{sec:methods_gasmass}).
After subtracting the best-fit two-component disk model from the CO(7--6) data, we detected stream-like residual emission that appears to bridge the radii of the two \ci\ disk components (Figure~\ref{fig:residual}). 
Non-axisymmetric gravitational torques can remove angular momentum from the gas and drive inflow toward the central few hundred parsecs \citep{HopkinsQuataert2010}, producing coherent non-circular structures such as spirals, streams, or arcs. Although the residual feature does not uniquely determine the gas flow geometry, its morphology is consistent with torque-driven inflow and suggests a plausible pathway by which the extended reservoir could supply material to the compact, highly excited nuclear disk.

The large intrinsic velocity dispersion ($\sigma_{\rm gas} = 277^{+16}_{-14}$~\kms) observed in the nuclear region is also naturally expected in this physical picture. 
Several processes could contribute: (i) a dense, geometrically thick nuclear disk in which turbulent and pressure support are dynamically important \citep{Wada2012}, (ii) feedback-driven turbulence powered by the radiative and mechanical output of the obscured AGN, including disk winds expected in super-critical accretion flows \citep{Jiang2014,Sadowski2014} and radiative feedback shaping the circumnuclear environment \citep{Ricci2017}, and (iii) shocks associated with angular-momentum transport and inflow from the extended reservoir \citep{HopkinsQuataert2010}. 
While the relative contributions of these processes cannot be disentangled with the present data, each is characteristic of a compact, heavily obscured nucleus undergoing rapid black-hole growth.

\section{Summary}
\label{sec:summary}

We have presented ultra-high-resolution ($0\farcs03 \approx 230$~pc) ALMA observations of the Hot DOG W2305$-$0039 at $z=3.111$.
By combining visibility-domain modeling of the \co(7--6), \co(11--10), \ci, and dust-continuum emission with forward modeling of the nuclear \co(11--10) position--velocity diagram, we obtained dynamical constraints on the circumnuclear ISM and the central black hole.
Our main results are as follows:

\begin{enumerate}
\item Visibility-plane modeling shows that the highly excited molecular component traced by \co(11--10) is extremely compact ($R_{\rm e}=173\pm30$~pc) and more concentrated than the lower-excitation \co(7--6) emission ($R_{\rm e}=259\pm58$~pc).
The 1.0~mm dust continuum is even more centrally concentrated ($R_{\rm e}=77\pm3$~pc), indicating that the dominant power source is confined to the inner $\sim$100~pc.

\item The \co(11--10)/\co(7--6) luminosity ratio rises steeply above unity within the central $\sim$500~pc.
This excitation cannot be reproduced by PDR models even for extreme UV radiation fields, whereas XDR models require intense X-ray irradiation on sub-kiloparsec scales, consistent with heating by a deeply embedded AGN co-spatial with the compact dust emission.

\item The nuclear \co(11--10) kinematics are dispersion-dominated, with a weak ordered component detectable along the major axis.
Forward modeling of the PV diagram yields a dynamical black-hole mass of
$\log(M_{\rm BH}/\Msun)=8.3^{+0.7}_{-0.6}$ and a large intrinsic gas velocity dispersion of
$\sigma_{\rm gas}=277^{+16}_{-14}$~\kms.

\item Combining $M_{\rm BH}$ with the independently inferred bolometric luminosity from infrared SED decomposition implies a super-Eddington accretion state, with $\log\lambda_{\rm Edd}=1.3^{+0.6}_{-0.7}$ ($\lambda_{\rm Edd} \gtrsim 4$).
For a fiducial radiative efficiency $\epsilon=0.1$, the implied accretion rate is $\dot{M}_{\rm acc}\simeq 90~\Msun~{\rm yr}^{-1}$.

\item The \ci\ emission reveals both a compact nuclear reservoir and a more extended disk, with inferred molecular gas masses of $(6.3\pm3.5)\times10^{9}~\Msun$ (compact) and $\gtrsim(3.1\pm1.1)\times10^{10}~\Msun$ (extended; a lower limit given uncertainties in excitation and abundance).
After subtracting the best-fit axisymmetric model, the \co(7--6) residuals show faint stream-like structure bridging the two components, suggesting a plausible pathway for inward gas transport that can sustain the compact, highly excited nuclear phase.
\end{enumerate}

Taken together, these results provide dynamical evidence that a hyperluminous, heavily obscured galaxy can host a compact, X-ray-heated molecular structure and a rapidly accreting black hole with $\lambda_{\rm Edd}\gg1$.
W2305$-$0039 therefore offers a physically motivated example of a short-lived evolutionary stage in which rapid black-hole growth proceeds while the nucleus remains deeply obscured by gas and dust, potentially preceding the emergence of optically luminous quasars.

Although extremely rare, luminous dust-obscured galaxies like W2305$-$0039 can be robustly detected at $z \simeq 3$--$4$; however, similarly luminous systems have not yet been found at $z \geq 6$. 
If the most rapid episodes of black-hole growth preferentially occur during such obscured phases, a substantial fraction of early accretion history may be missed by existing quasar surveys. 
Systems like W2305$-$0039 may therefore trace an important yet largely hidden channel of black-hole growth in the early Universe. 
Identifying such progenitors is challenging, because extreme dust obscuration requires photometric coverage at wavelengths longer than $\sim$3~$\mu$m to establish their red colors.
Wide-area surveys limited to $\leq$2~$\mu$m, such as LSST, Euclid, or Roman, lack the wavelength leverage to identify extreme dust-obscured galaxies at $z \geq 6$. 
Progress will likely require wide-area surveys at mid-infrared and submillimeter wavelengths reaching sensitivities well beyond those of \emph{WISE}. 
Forthcoming facilities combining high sensitivity with large sky coverage, including next-generation infrared space missions (e.g., GREX-PLUS; \citealt{2024SPIE13092E..0YI}) and 50-m-class single-dish submillimeter telescopes (e.g., AtLAST/LST; \citealt{2025AA...694A.142M}; \citealt{2020SPIE11453E..0NK}), offer a promising route to uncovering these elusive populations at $z \geq 6$. 

Once identified, ultra-high-resolution observations of compact, highly excited molecular gas will provide a powerful dynamical probe of black-hole growth in such systems. 
By resolving high-$J$ CO emission on sub-kiloparsec scales, ALMA can already deliver dynamical constraints on black-hole masses even in optically opaque, turbulence-dominated systems, opening a new window on the most obscured phases of supermassive black-hole assembly. 
Looking ahead, proposed configurations that extend ALMA's maximum baselines beyond the current $\sim$16~km to $\sim$30--50~km would push the angular resolution in Bands~6--7 to $\lesssim 0\farcs01$, corresponding to tens of parsecs at $z>6$ \citep{2024arXiv240907570O}. 
Because high-$J$ CO emission in dusty systems can be exceptionally compact, it should remain detectable on such long baselines, enabling kinematic measurements on scales approaching the black-hole sphere of influence in hosts of $M_{\rm BH}\sim10^{9}\,\Msun$ black holes. 
Resolving this scale would substantially reduce the dependence of dynamical mass measurements on assumptions about the host-galaxy mass distribution and would provide a stringent benchmark for models of rapid black-hole assembly in the early Universe.
Combined with wide-area mid-infrared and submillimeter surveys capable of identifying the most heavily obscured progenitors, such observations would directly connect obscured, rapid-growth phases to the emergence of optically luminous quasars at the highest redshifts.

\begin{acknowledgments}
The author thanks the anonymous referee for the careful reading of the manuscript and for the constructive comments that improved this paper. 
This paper makes use of the following ALMA data: ADS/JAO.ALMA\#2024.1.01175.S, ADS/JAO.ALMA\#2022.1.00353.S, and ADS/JAO.ALMA\#2021.1.00168.S.
ALMA is a partnership of ESO (representing its member states), NSF (USA), and NINS (Japan), together with NRC (Canada), NSTC and ASIAA (Taiwan), and KASI (Republic of Korea), in cooperation with the Republic of Chile.
The Joint ALMA Observatory is operated by ESO, AUI/NRAO, and NAOJ.
The author used ChatGPT (OpenAI) and Claude (Anthropic) for assistance with language
editing and code refinement. The author takes full responsibility for the content of this manuscript.
K.T. was supported by the ALMA Japan Research Grant of NAOJ ALMA Project, NAOJ-ALMA-386.
K.T. acknowledges support from JSPS KAKENHI Grant Number JP 23K03466.
\end{acknowledgments}

\facilities{ALMA}

\software{CASA \citep{CASA2022}, UVMULTIFIT \citep{MartiVidal2014}, galaxySLED \citep{Esposito2024}, KinMS \citep{Davis2013}, emcee \citep{ForemanMackey2013}}

\bibliographystyle{apj}
\bibliography{ref}

\end{document}